\documentclass[final,5p,times,twocolumn]{elsarticle}
\journal{Phys. Lett. B}

\usepackage{graphicx,color}
\usepackage{amsmath,amssymb}

\newcommand{\eq}[1]{\begin{equation}#1\end{equation}}
\newcommand{\eqmulti}[1]{\begin{equation}\begin{split}#1\end{split}\end{equation}}

\newcommand{\bra}[1]{\ensuremath{\langle{#1}|\,}}

\newcommand{\ket}[1]{\ensuremath{\,|{#1}\rangle}}

\newcommand{\braket}[2]{\ensuremath{\langle{#1}|{#2}\rangle}}

\newcommand{\matrixe}[3]{\ensuremath{\langle{#1}|\,{#2}\,|{#3}\rangle}}

\newcommand{\op}[1]{\ensuremath{#1}}

\newcommand{\HO}{\ensuremath{\op{H}}}

\newcommand{\TO}{\ensuremath{\op{T}}}

\newcommand{\VO}{\ensuremath{\op{V}}}
\newcommand{\WO}{\ensuremath{\op{W}}}

\newcommand{\cm}{\ensuremath{\textrm{cm}}}
\newcommand{\elem}[2]{\ensuremath{{}^{#2}\text{#1}}}

\newcommand{\symboldiamond}[1][black]{{\color{#1}$\blacklozenge$}}
\newcommand{\symboltriangle}[1][black]{{\color{#1}$\blacktriangle$}}

\newcommand{\symbolcircle}[1][black]{{\color{#1}$\bullet$}}

\definecolor{FGViolet}{rgb}{0.61,0.32,0.61}
\definecolor{FGDarkBlue}{rgb}{0,0,0.6}
\definecolor{FGBlue}{rgb}{0,0,0.8}
\definecolor{FGLightBlue}{rgb}{0.2, 0.6, 0.8}
\definecolor{FGGreen}{rgb}{0.2,0.7,0.2}
\definecolor{FGLightGreen}{rgb}{0.4,1,0.4}
\definecolor{FGYellow}{rgb}{1,0.95,0}
\definecolor{FGOrange}{rgb}{0.95,0.5,0.1}
\definecolor{FGRed}{rgb}{0.8,0,0}
\definecolor{FGWhite}{rgb}{1,1,1}
\definecolor{FGLightGray}{rgb}{0.8,0.8,0.8}
\definecolor{FGGray}{rgb}{0.5,0.5,0.5}
\definecolor{FGDarkGray}{rgb}{0.3,0.3,0.3}
\definecolor{FGBlack}{rgb}{0,0,0}

\begin{document}

\begin{frontmatter}

\title{Pad\'e-resummed high-order perturbation theory for nuclear structure calculations}

\author[tud]{Robert Roth}
\ead{robert.roth@physik.tu-darmstadt.de}

\author[tud]{Joachim Langhammer}

\address[tud]{Institut f\"ur Kernphysik, Technische Universit\"at Darmstadt,
64289 Darmstadt, Germany}

\begin{abstract}  
We apply high-order many-body perturbation theory for the calculation of ground-state energies of closed-shell nuclei using realistic nuclear interactions. Using a simple recursive formulation, we compute the perturbative energy contributions up to 30th order and compare to exact no-core shell model calculations for the same model space and Hamiltonian. Generally, finite partial sums of this perturbation series do not show convergence with increasing order, but tend to diverge exponentially. Nevertheless, through a simple resummation via Pad\'e approximants it is possible to extract rapidly converging and highly accurate results for the ground state energy.

\end{abstract}


\end{frontmatter}

\section{Introduction}

The treatment of the nuclear many-body problem is a central and long-standing issue in nuclear structure theory. Ideally, we would like to solve the many-body problem \emph{ab initio}, i.e., starting from a given nuclear Hamiltonian without any conceptual approximations. With the advent of high-precision nuclear potentials that are based systematically on Quantum Chromodynamics (QCD) through chiral effective field theory \cite{EnMa03,EpNo02}, the demand for exact \emph{ab initio} solutions of the nuclear many-body problem has grown. Only these schemes establish a rigorous and quantitative connection between nuclear structure observables and the underlying QCD input. 

The no-core shell model (NCSM) is one of the most universal exact \emph{ab initio} methods, which gives access to all aspects of nuclear structure \cite{NaQu09,MaVa09,RoNa07}. Other methods, are either restricted to certain classes of Hamiltonians, like the Green's Function Monte Carlo approach \cite{PiWi04}, or they are limited to certain nuclei and observables, like the coupled-cluster approach \cite{HaPa08}. All of them are computationally demanding, which leads to a severe limitation regarding the number of nucleons that can be handled.

Therefore, approximate many-body schemes using the same Hamiltonians, i.e. approximate \emph{ab initio} methods, also provide indispensable information. In particular approaches that use controlled and systematically improvable approximations are of great practical importance. In this category, many-body perturbation theory (MBPT) is one of the most powerful and widely used methods. On the one hand, the evaluation of low-orders of perturbation theory is computationally simple and can be done for the whole nuclear mass range \cite{RoPa06,CoCo06,CoCo07,StSt01} as well as for infinite nuclear matter \cite{BoSc05}. On the other hand, it is deemed systematically improvable, either by extending the MBPT calculations order-by-order or by using infinite partial summations, like ladder- or ring-type summations \cite{StJo02,BaPa06,SiHo09}. However, the accuracy of low-order perturbative estimates, e.g. for ground-state energies, or possible extensions of the MBPT series to higher orders and the resulting convergence pattern are rarely, if ever, addressed in the nuclear structure context.

In this paper, we apply MBPT for the calculation of the ground state energy of several closed-shell nuclei. We extend the order-by-order calculation of the perturbative energy contributions up to 30th order, study the convergence behavior, and compare to exact NCSM calculations for the same Hamiltonian and model space. We introduce Pad\'e approximants as a highly efficient tool for the resummation of the divergent power-series of MBPT into a rapidly converging series and demonstrate their accuracy for the description of ground-state energies.

\section{Many-body perturbation theory}
\label{sec:mbpt}

\subsection{Formalism}

We aim at a perturbative expansion of the many-nucleon Schr\"odinger equation 
\eq{ \label{eq:schroedinger}
 \HO \ket{\Psi_n} = E_n \ket{\Psi_n}
}
for the translational invariant nuclear Hamiltonian $\HO = \TO-\TO_{\cm}+\VO$, where we assume $\VO$ to be a two-body interaction for simplicity. In a first step we have to chose the unperturbed basis, which in turn defines the unperturbed Hamiltonian. From the practical point of view, a basis of Slater-determinants constructed from a set of single-particle states is most convenient. The underlying single-particle basis will typically be a Hartree-Fock or a harmonic oscillator basis---for simplicity we assume the latter. The unperturbed Hamiltonian $\HO_0$ is a one-body operator containing the kinetic energy $\TO$ and a harmonic oscillator potential. The unperturbed Slater determinants $\ket{\Phi_n}$ fulfill the eigenvalue relation  
\eq{
  \HO_0 \ket{\Phi_n} = \epsilon_n \ket{\Phi_n}
}
with eigenvalues $\epsilon_n$ being the sum of the single-particle energies of the occupied states.  After the unperturbed Hamiltonian is fixed, the perturbation is defined through $\WO = \HO-\HO_0$. This partitioning leads to the M\o{}ller-Plesset formulation of MBPT and obviously other partitionings of the Hamiltonian are possible \cite{SzOs96,Roth09}. For ease of presentation, we assume that the unperturbed state corresponding to the eigenstate we are interested in is non-degenerate, as it is the case for the ground state of closed shell nuclei. In the case of degeneracy, as e.g. for the excited states of closed shell nuclei, one would have to diagonalize the full Hamiltonian in the degenerate subspace and pick the eigenstates with the desired quantum numbers as unperturbed states.

The standard Rayleigh-Schr\"odinger perturbation series can now be constructed based on a Hamiltonian (using the notation from Ref. \cite{Roth09})
\eq{ \label{eq:hamiltonian_lambda}
  \HO(\lambda) = \HO_0 + \lambda\, \WO
}
containing an auxiliary expansion parameter $\lambda$ that continuously connects the unperturbed Hamiltonian $\HO_0=\HO(\lambda=0)$ with the full Hamiltonian $\HO = \HO(\lambda=1)$. The energy eigenvalues $E_n(\lambda)$ and the corresponding eigenvectors $\ket{\Psi_n(\lambda)}$ of $\HO(\lambda)$ are formulated as a power series in $\lambda$
\eqmulti{ \label{eq:pseries_ansatz}
  E_n(\lambda) 
  & = E_n^{(0)} + \lambda E_n^{(1)} + \lambda^2 E_n^{(2)} + \dots\,, \\
  \ket{\Psi_n(\lambda)}
  & = \ket{\Psi_n^{(0)}} + \lambda \ket{\Psi_n^{(1)}} + \lambda^2 \ket{\Psi_n^{(2)}} + \dots \;.
}
In the absence of degeneracy the lowest-order contributions are simply given by the unperturbed quantities, i.e.,
\eq{
  E_n^{(0)} = \epsilon_n \;,\qquad
  \ket{\Psi_n^{(0)}} = \ket{\Phi_n} \;.
}

Inserting the Hamiltonian \eqref{eq:hamiltonian_lambda} and the power series \eqref{eq:pseries_ansatz} into the Schr\"odinger equation \eqref{eq:schroedinger} leads to the fundamental equation 
\eqmulti{ \label{eq:fundamental}
  &\HO_0 \ket{\Psi_n^{(0)}} + \sum_{p=1}^{\infty} \lambda^p
    \big( \WO \ket{\Psi_n^{(p-1)}} + \HO_0 \ket{\Psi_n^{(p)}} \big) \\
  &\qquad\qquad= E_n^{(0)} \ket{\Psi_n^{(0)}} + \sum_{p=1}^{\infty} \lambda^p
    \bigg(\sum_{j=0}^{p} E_n^{(j)} \ket{\Psi_n^{(p-j)}} \bigg) \;.
}
Assuming that the unperturbed states form an orthonormal basis and using the intermediate normalization $\braket{\Psi_n^{(0)}}{\Psi_n(\lambda)}=1$ we obtain $\braket{\Psi_n^{(0)}}{\Psi_n^{(p)}}=0$ for $p>0$, which allows us to project-out all required information on the individual contributions in the power series. By multiplying Eq. \eqref{eq:fundamental} with $\bra{\Psi_n^{(0)}}$ and matching same orders of $\lambda$ on both sides, we immediately obtain a simple expression for the $p$th-order energy contribution
\eq{ \label{eq:pt_energy1}
  E_n^{(p)} = \matrixe{\Psi_n^{(0)}}{\WO}{\Psi_n^{(p-1)}} \;.
}
By multiplying Eq. \eqref{eq:fundamental} with $\bra{\Psi_m^{(0)}}$ with $m\neq n$ and matching $\lambda$-orders, we obtain an expression for the amplitudes 
\eqmulti{ \label{eq:pt_amplitude1}
  &C_{n,m}^{(p)} = \braket{ \Psi_m^{(0)} }{ \Psi_n^{(p)} } = \\
  &= \frac{1}{E_n^{(0)} - E_m^{(0)}}
  \bigg( \matrixe{\Psi_m^{(0)}}{\WO}{\Psi_n^{(p-1)}} - \sum_{j=1}^p E_n^{(j)} \braket{\Psi_m^{(0)}}{\Psi_n^{(p-j)}} \bigg)
}  
which characterize the perturbative corrections to the eigenstates $\ket{\Psi_n^{(p)}}$ expanded in the unperturbed basis
\eq{
  \ket{\Psi_n^{(p)}} = \sum_{m} C_{n,m}^{(p)} \ket{\Psi_m^{(0)}} 
}
with $C_{n,n}^{(p)}=0$ for $p>0$ and $C_{n,m}^{(0)}=\delta_{n,m}$.

We can cast Eqs. \eqref{eq:pt_energy1} and \eqref{eq:pt_amplitude1} into a more transparent form by systematically introducing the amplitudes $C_{n,m}^{(p)}$ and formulating all matrix elements in terms of the unperturbed states. For the $p$th-order energy contribution we obtain 
\eq{ \label{eq:pt_energy}
  E_n^{(p)}
  = \sum_{m} \matrixe{\Phi_n}{\WO}{\Phi_m}\; C_{n,m}^{(p-1)} \;.
}
Similarly we obtain for the $p$th-order amplitudes
\eq{ \label{eq:pt_amplitude}
  C_{n,m}^{(p)} 
  = \frac{1}{\epsilon_n - \epsilon_m}
  \bigg( \sum_{m'} \matrixe{\Phi_m}{\WO}{\Phi_{m'}} C_{n,m'}^{(p-1)} - \sum_{j=1}^p E_n^{(j)}  C_{n,m}^{(p-j)} \bigg) .
}
Together with $C_{n,m}^{(0)}=\delta_{n,m}$ and $E_n^{(0)}=\epsilon_n$ these relations form a recursive set of equations which uniquely determines the perturbative corrections for all energies and states to all orders.

Usually one would use these general expressions to derive explicit formulae for the lowest-order corrections. The matrix elements of the perturbation in the unperturbed Slater-determinant states can be evaluated explicitly and the summations over the many-body basis set can be replaced by summations over single-particle states. In this way we would recover the standard expressions for, e.g., the second- and third-order energy corrections \cite{RoPa06,StSt01,SzOs96}.

\subsection{Evaluation to high orders}
  
When attempting to evaluate the perturbative corrections beyond third- or forth-order the explicit formulae for the energy corrections become impractical because of the large number of nested summations. A much more elegant way to evaluate high-order contributions makes use of the recursive structure of Eqs. \eqref{eq:pt_energy} and \eqref{eq:pt_amplitude}. The only ingredients needed are the many-body matrix elements of the full Hamiltonian $\HO$ with respect to the unperturbed basis $\ket{\Phi_n}$. Starting from the zeroth-order coefficients $C_{n,m}^{(0)}=\delta_{n,m}$ we can readily evaluate the first-order energy contribution $E_n^{(1)}$ from \eqref{eq:pt_energy}. This in turn allows us to compute the first-order coefficients $C_{n,m}^{(1)}$ via \eqref{eq:pt_amplitude}. Generally, for the evaluation of the energy contribution $E_n^{(p)}$ only the coefficients $C_{n,m}^{(p-1)}$ of the previous order are required. For the evaluation of the coefficients $C_{n,m}^{(p)}$ all energy contributions up to order $p$ and all coefficients up to order $(p-1)$ need to be known. 

Technically, the recursive evaluation of the perturbation series bears some resemblance to the Lanczos algorithm for the iterative solution of the eigenvalue problem for a few extremal eigenvalues as it is used in the NCSM. The most significant operation is a matrix-vector multiplication of the Hamiltonian matrix with the coefficient vector from the previous order, which constitutes the first term in the evaluation of the coefficients \eqref{eq:pt_amplitude}. Because the second term in \eqref{eq:pt_amplitude} involves the coefficient vectors from all previous orders, we store them for simplicity. These computational elements are the same as for a simple Lanczos algorithm in the NCSM or in corresponding configuration interaction (CI) approaches, therefore, an implementation of high-order MBPT using NCSM or CI technologies is straight forward. However, since the computational elements are the same, so are the computational limitations: This direct implementation of high-order MBPT is limited to the same model spaces and particle numbers as the full NCSM. This is not a concern for the present study, but for an application of MBPT beyond the domain of the NCSM one has to resort to other evaluation schemes.

\subsection{Applications: \elem{He}{4}, \elem{O}{16} and \elem{Ca}{40}}

\begin{figure*}[t]
\centering\includegraphics[width=1\textwidth]{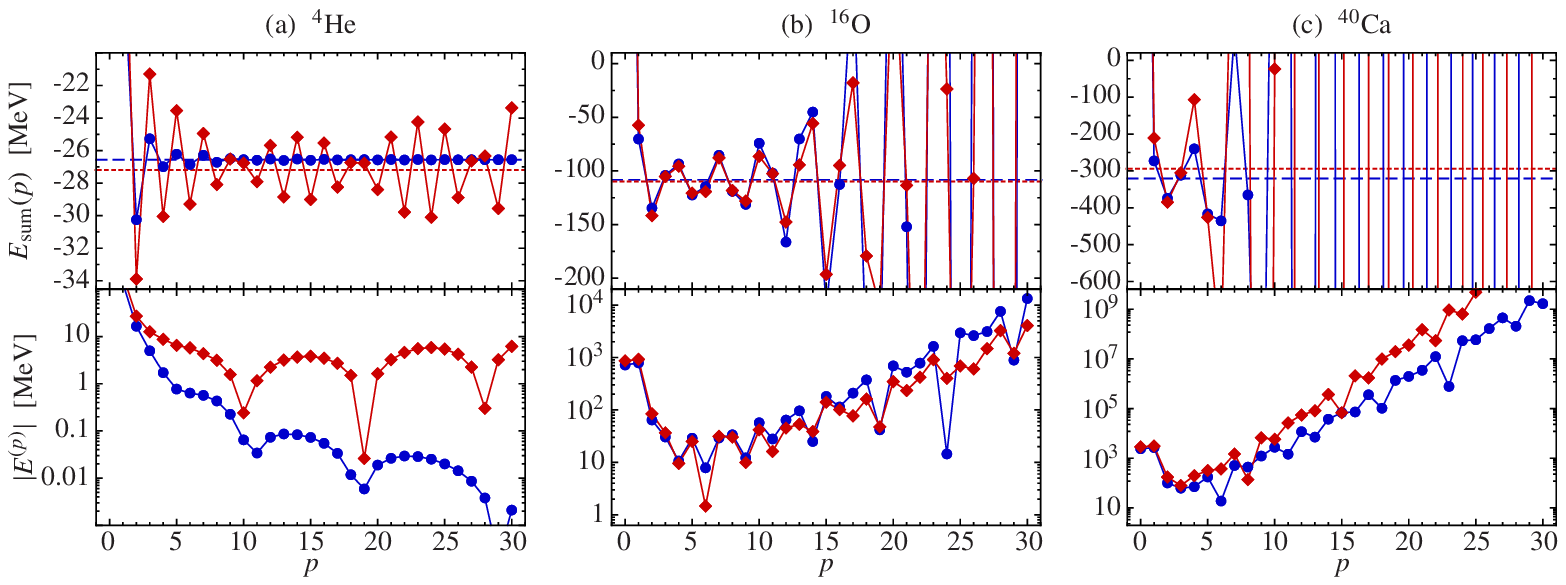}
\caption{Contributions to the ground-state energies in MBPT up to 30th order for different nuclei and model spaces: (a) \elem{He}{4} in a $12\hbar\Omega$ model space with $\hbar\Omega=20$ MeV (\symbolcircle[FGBlue]) and $32$ MeV (\symboldiamond[FGRed]); (b) \elem{O}{16} in $6\hbar\Omega$ with $\hbar\Omega=20$ MeV (\symbolcircle[FGBlue]) and $24$ MeV (\symboldiamond[FGRed]); (c) \elem{Ca}{40} in $4\hbar\Omega$ with $\hbar\Omega=20$ MeV (\symbolcircle[FGBlue]) and $24$ MeV (\symboldiamond[FGRed]). The upper panels depict the partial sum $E_{\text{sum}}(p)$ defined in Eq. \eqref{eq:runningsum} as function of the largest order $p$, the lower panels the modulus of the individual contributions $E^{(p)}$ on a logarithmic scale. The dashed horizontal lines in the upper panels indicate the NCSM ground-state energies for the respective nuclei and model spaces.}
\label{fig:series}
\end{figure*}

As examples for a direct application of high-order MBPT we consider the ground-state energies of \elem{He}{4}, \elem{O}{16}, and \elem{Ca}{40}. Throughout this work we use an intrinsic Hamiltonian with a soft two-nucleon interaction that is derived from the chiral N3LO potential \cite{EnMa03} via a Similarity Renormalization Group (SRG) transformation \cite{BoFu07,HeRo07,RoRe08}. The final flow parameter for the SRG evolution of the interaction is $\alpha=0.02\,\text{fm}^4$ which corresponds to a momentum scale of $\Lambda=2.66\,\text{fm}^{-1}$. This choice for the flow parameters leads to a unitarily transformed interaction which is sufficiently soft to warrant excellent convergence properties with respect to model space size in the NCSM but at the same time yields ground-state energies which are in reasonable correspondence with experiment in the mass-range considered here. To allow for a direct comparison with exact NCSM calculations for the same Hamiltonian and the same model space, we use an $N_{\max}\hbar\Omega$ model space also for the MBPT calculations. We have confirmed, however, that all conclusions regarding the performance and limitations of the MBPT approach do not depend on this particular choice.

In Fig.~\ref{fig:series} we summarize the results of an order-by-order MBPT calculation up to 30th order for the ground state energy of the three nuclei. For $\elem{He}{4}$ the calculations were performed in a $12\hbar\Omega$ model space, for \elem{O}{16} in $6\hbar\Omega$, and for \elem{Ca}{40} in $4\hbar\Omega$, each with two different oscillator frequencies $\hbar\Omega$. The partial sum of the perturbative energy contributions up to order $p$, 
\eq{ \label{eq:runningsum}
  E_{\text{sum}}(p) 
  = \sum_{p'=0}^{p} E^{(p')} \;,
}
is depicted in upper row and the modulus of the individual $p$th-order contributions, $|E^{(p)}|$, on a logarithmic scale in the lower row. Here and in the following we omit the index $n=0$ for convenience.

Already the first glance at Fig.~\ref{fig:series} reveals a fundamental problem with the convergence behavior of the perturbation series. For \elem{He}{4}, as depicted in Fig.~\ref{fig:series}(a), we observe two different patterns depending on the oscillator frequency. For $\hbar\Omega=20$ MeV the partial sum $E_{\text{sum}}(p)$ shows an alternating behavior with a systematically decreasing amplitude. Beyond 10th order one might consider the perturbation series converged and the resulting energy is in excellent agreement with the result of an NCSM calculation with the same Hamiltonian in the same model space. However, a change of the oscillator frequency destroys this picture. For $\hbar\Omega=32$ MeV, where the NCSM provides a lower ground-state energy, we again observe an alternating sequence of energy contributions $E^{(p)}$, but this time without any sign of convergence. The absolute value of the individual energy corrections does not decrease with increasing order, it even shows a slightly increasing trend. Hence even in the simplest case, the \elem{He}{4} ground state, the convergence of the perturbation series is not guaranteed. 

The situation is even more dramatic for \elem{O}{16} or \elem{Ca}{40} as depicted in Fig.~\ref{fig:series}(b) and (c). In all cases the size of the perturbative energy contributions $|E^{(p)}|$ grows \emph{exponentially} with $p$. The partial sum $E_{\text{sum}}(p)$ exhibits a strong oscillatory behavior with increasing amplitude. Only the lower orders, typically up to 10th order for \elem{O}{16} and up to 5th order for \elem{Ca}{40}, lead to binding energies in a physically meaningful energy range. At the 30th order the perturbative contributions are in the order of $10^4$ MeV for \elem{O}{16} and $10^9$ MeV for \elem{Ca}{40}---this is beyond any physical energy scale present in the nuclear many-body problem. We were not able to find a convergent scenario by varying the oscillator frequency or the model space size or truncation for these nuclei.

The explosion of the perturbative corrections beyond any meaningful energy scale suggests a principal mathematical problem in the representation of the energy eigenvalue $E(\lambda)$ as a partial sum of a simple power series \eqref{eq:pseries_ansatz} .

\section{Pad\'e approximants}
\label{sec:pade}

\subsection{Formalism}

Prompted by the drastic failure of a partial sum of a simple power series to describe the energy $E(\lambda)$ at the physical point $\lambda=1$ one might consider more general expansions of this function. A next step would be an expansion of the energy $E(\lambda)$ in terms of a rational function composed of separate power series for numerator and denominator   
\eq{
  E(\lambda) 
  = \frac{A(\lambda)}{B(\lambda)}
  = \frac{a_0 + a_1 \lambda + a_2 \lambda^2 + \dots}{b_0 + b_1 \lambda + b_2 \lambda^2 + \dots} \;.
}
Obviously we will not attempt to re-derive perturbation theory for this type of expansion. The above is useful only, if we could use the information contained in the standard MBPT energy contributions $E^{(p)}$ to construct this rational expansion.

Exactly this is achieved through the Pad\'e approximants \cite{Bake65,BaGr96}. Given a power series \eqref{eq:pseries_ansatz} of the function $E(\lambda)$, then the Pad\'e approximant
\eq{
  [M/N](\lambda)
  = \frac{a_0 + a_1 \lambda + a_2 \lambda^2 + \dots + a_M \lambda^M}{b_0 + b_1 \lambda + b_2 \lambda^2 + \dots + b_N \lambda^N}
}
with numerator being a polynomial of order $M$ and the denominator a polynomial of order $N$ is constructed such that its Taylor expansion reproduces the first $M+N$ orders of the initial power series, i.e.
\eq{
  E(\lambda) = [M/N](\lambda) + O(\lambda^{M+N+1}) \;.
}
From this definition one can immediately construct a coupled system of equations that determines the coefficients $a_n$ and $b_m$ of the Pad\'e approximant from a given set of $E^{(p)}$ with $p=0,...,N+M$. An alternative and more elegant form \cite{Bake65,BaGr96} relates the Pad\'e approximants to determinants of two $(N+1)\times(N+1)$ matrices containing directly the power-series coefficients $E^{(p)}$ 
\eq{ \label{eq:padeapprox}
[M/N](\lambda) = \small
\frac{\left|
\begin{array}{cccc}
E^{(M-N+1)} & E^{(M-N+2)} & \cdots & E^{(M+1)}\\
E^{(M-N+2)} & E^{(M-N+3)} & \cdots & E^{(M+2)}\\
\vdots & \vdots & \ddots & \vdots\\
E^{(M)} & E^{(M+1)} & \cdots & E^{(M+N)}\\
\sum\limits_{i=0}^{M-N} E^{(i)} \lambda^{N+i} & \sum\limits_{i=0}^{M-N+1} E^{(i)}\lambda^{N+i-1} & \cdots & \sum\limits_{i=0}^{M} E^{(i)}\lambda^{i}
\end{array}
\right|}
{\left|
\begin{array}{ccccc}
E^{(M-N+1)} & E^{(M-N+2)} & \cdots & E^{(M+1)}\\
E^{(M-N+2)} & E^{(M-N+3)} & \cdots & E^{(M+2)}\\
\vdots & \vdots & \ddots & \vdots\\
E^{(M)} & E^{(M+1)} & \cdots &  E^{(M+N)}\\
\lambda^{N} & \lambda^{N-1} & \cdots & 1
\end{array}
\right|}\,,
}
where we set $E^{(p)}\equiv 0$ for $p<0$. We will use this form to evaluate various Pad\'e approximants in the following.

Before considering numerical results, we should like to mention a few formal properties of the Pad\'e approximants that are of importance for the present application. The mathematical foundation for using Pad\'e approximants for our purpose in the first place is provided by the Pad\'e conjecture (simplified) \cite{Bake65,BaGr96}: Let $E(\lambda)$ be a continuous function for $|\lambda|\leq1$, then there is an infinite subsequence of diagonal Pad\'e approximants $[N/N](\lambda)$ that for $N\to\infty$ converges locally uniformly against $E(\lambda)$ for $|\lambda|\leq1$. For our application the continuity requirements for the function $E(\lambda)$ are always fulfilled, thus we expect the diagonal Pad\'e approximants to show a convergence behavior---unlike the simple power series.

Additionally the Pad\'e approximants have a number of specific properties that would be extremely valuable in the present context.
If the power series expansion of $E(\lambda)$ is a Stieltjes series, then the Pad\'e approximants fulfill the condition
\eq{ \label{eq:inequality}
  [M/M](\lambda) \geq E(\lambda) \geq [M-1/M](\lambda)
}
for $\lambda\geq0$ as well as a whole set of related inequalities \cite{Bake65,BaGr96}. Thus the diagonal and the super-diagonal Pad\'e approximants provide upper and lower bounds for the full energy $E(\lambda)$, respectively. Furthermore, these bounds improve monotonically with increasing order $M$ of the Pad\'e approximant. Unfortunately, it turns out that the power series we start from is not a Stieltjes series in general.

\subsection{Applications: \elem{He}{4}, \elem{O}{16} and \elem{Ca}{40}}

\begin{figure*}[t]
\centering\includegraphics[width=1\textwidth]{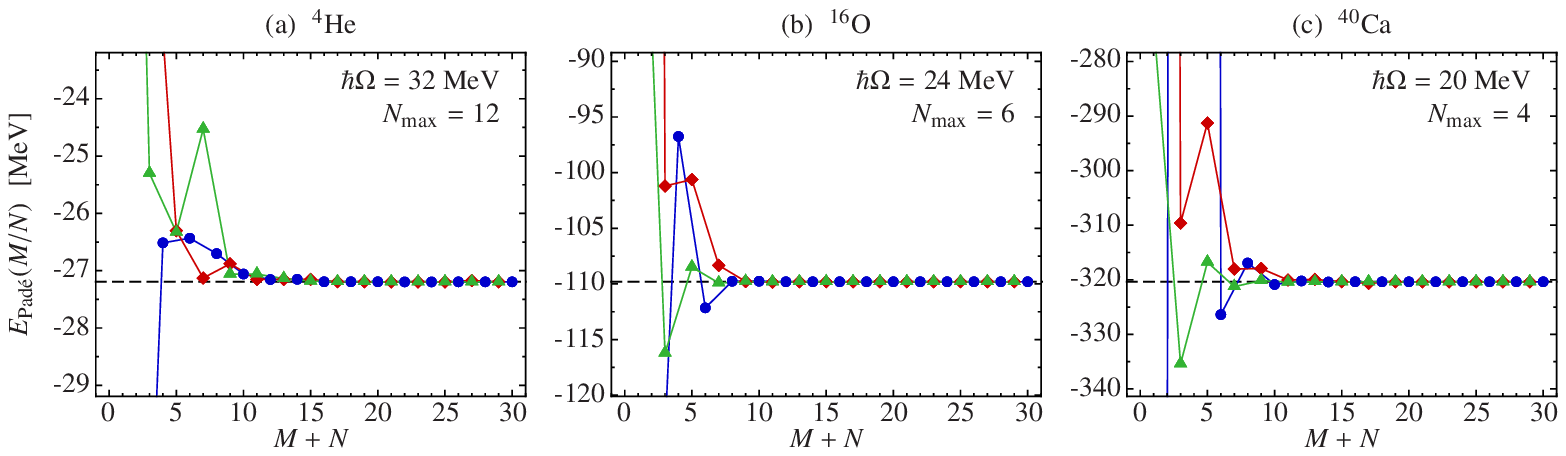}
\caption{Pad\'e approximants for the ground-state energies of (a) \elem{He}{4}, (b) \elem{O}{16}, and (c) \elem{Ca}{40} as function of the summed order $M+N$. The different symbols represent the diagonal approximants $E_{\text{Pad\'e}}(M/M)$ (\symbolcircle[FGBlue]), the super-diagonal approximants $E_{\text{Pad\'e}}(M-1/M)$ (\symboldiamond[FGRed]), and the sub-diagonal approximants $E_{\text{Pad\'e}}(M/M-1)$ (\symboltriangle[FGGreen]). The model space size $N_{\max}$ and the oscillator frequency $\hbar\Omega$ is quoted in the individual panels. The dashed horizontal lines indicate the NCSM ground-state energies for the respective nuclei and model spaces.} 
\label{fig:pade}
\end{figure*}

Using the results of the order-by-order calculation of the energy corrections $E^{(p)}$ up to 30th order of MBPT we can construct all Pad\'e approximants with $N+M \leq 30$ from Eq. \eqref{eq:padeapprox}. Evaluating the approximant at $\lambda=1$ yields an estimate for the ground-state energy of the perturbed system
\eq{
  E_{\text{Pad\'e}}(M/N) = [M/N](\lambda=1) \;.
}
We will focus on the diagonal Pad\'e approximant, $E_{\text{Pad\'e}}(M/M)$, and the super- and sub-diagonal approximants, $E_{\text{Pad\'e}}(M-1/M)$ and $E_{\text{Pad\'e}}(M/M-1)$, respectively, because of the convergence and boundary theorems available for those. 

A collection of all diagonal as well as sub- and super-diagonal approximants with $N+M \leq 30$ for \elem{He}{4}, \elem{O}{16}, and \elem{Ca}{40} using the oscillator frequencies that yield the lowest ground-state energy is provided in Fig. \ref{fig:pade}. The first remarkable observation is that the Pad\'e approximants converge very quickly for sufficiently large order---we have observed this behavior in all cases we considered. For $M+N\gtrsim10$ essentially all Pad\'e approximants provide the same ground-state energy. We emphasize that the input for the construction of those Pad\'e approximants are the exponentially diverging coefficients from the power-series formulation of MBPT discussed in Fig. \ref{fig:series}. The Pad\'e resummation of these coefficients efficiently regularizes these divergencies and leads to exceptionally stable results for all orders $M+N\gtrsim10$.

The second important observation results from the comparison of the converged Pad\'e approximants\footnote{Here, the term convergence refers solely to the convergence with respect to the order $M+N$ and not to convergence with respect to the model-space size $N_{\max}$, which is a separate issue.} with the exact energy eigenvalue obtained from a solution of the matrix eigenvalue problem for the Hamiltonian in the same model space---i.e., from the corresponding NCSM calculation---as indicated by the dashed horizontal line in Fig. \ref{fig:pade}. The converged Pad\'e approximants \emph{exactly} reproduce the corresponding energy eigenvalues, i.e., Pad\'e resummed perturbation theory and the exact solution of the eigenvalue problem become equivalent.

\begin{table}[t]
\caption{Large-scale MBPT results for the ground-state energies of \elem{He}{4}, \elem{O}{16}, and \elem{Ca}{40}. Shown are the exact NCSM energies for the respective model space and the deviations of the partial sums, $\Delta E_{\text{sum}}(p) = E_{\text{sum}}(p)-E_{\text{NCSM}}$, as well as the deviations of various Pad\'e approximants, $\Delta E_{\text{Pad\'e}}(M/N) = E_{\text{Pad\'e}}(M/N) - E_{\text{NCSM}}$. All energies are given in units of MeV.}
\label{tab:comparison}
\footnotesize 
\begin{tabular}{@{\extracolsep{-6pt}} l r r r r r r}
\hline
 &  \multicolumn{2}{c}{\elem{He}{4}}  &  \multicolumn{2}{c}{\elem{O}{16}}  &  \multicolumn{2}{c}{\elem{Ca}{40}}  \\ 
$N_{\max}$           &  12  &  12  &  6   &  6   &  4   &  4   \\
$\hbar\Omega$ [MeV]  &  20  &  32  &  20  &  24  &  20  &  24  \\
\hline
$E_{\text{NCSM}}$    & -26.561 & -27.194 & -108.33 & -109.81 & -320.37 &  -294.19 \\
\hline
$\Delta E_{\text{sum}}(1)$  & +12.865 & +20.320 & +38.05 & +52.39 & +47.62 & +83.50 \\
$\Delta E_{\text{sum}}(2)$  & -3.691 & -6.695 & -26.33 & -31.92 &  -53.68 &  -90.55 \\
$\Delta E_{\text{sum}}(3)$  & +1.288 & +5.896 &  +4.21 &  +4.64 &   +8.83 &  -11.00 \\
$\Delta E_{\text{sum}}(4)$  & -0.429 & -2.853 & +14.84 & +14.24 &  +80.27 & +187.51 \\
$\Delta E_{\text{sum}}(5)$  & +0.341 & +3.651 & -14.13 & -10.88 &  -96.18 & -131.81 \\
$\Delta E_{\text{sum}}(6)$  & -0.293 & -2.095 &  -6.27 &  -9.41 & -115.05 & -503.15 \\
$\Delta E_{\text{sum}}(7)$  & +0.275 & +2.247 & +22.93 & +21.98 & +395.80 & +985.80 \\
$\Delta E_{\text{sum}}(8)$  & -0.156 & -0.896 & -10.68 &  -8.42 &  -44.78 & +1124.02 \\
$\Delta E_{\text{sum}}(9)$  & +0.070 & +0.674 & -22.87 & -18.37 & -1270.07 & -5523.29 \\
$\Delta E_{\text{sum}}(10)$ & +0.005 & +0.433 & +34.16 & +23.47 & +1500.66 & +270.84 \\
\hline 
$\Delta E_{\text{Pad\'e}}(1/1)$    & -6.031 & -10.941 & -32.04 & -40.39 &   -57.68 & -100.93 \\
$\Delta E_{\text{Pad\'e}}(1/2)$    & +2.274 &  +7.759 &  +6.94 &  +8.61 &   +10.76 &   -5.94 \\
$\Delta E_{\text{Pad\'e}}(2/1)$    & +0.136 &  +1.894 &  -5.62 &  -6.41 &   -15.03 &  -35.95 \\
$\Delta E_{\text{Pad\'e}}(2/2)$    & +0.009 &  +0.680 & +14.93 & +13.06 & +3115.69 & -193.45 \\
$\Delta E_{\text{Pad\'e}}(2/3)$    & +0.108 &  +0.892 &  +8.19 &  +9.19 &   +29.10 &  +61.98 \\
$\Delta E_{\text{Pad\'e}}(3/2)$    & +0.066 &  +0.865 &  +1.01 &  +1.32 &    +3.61 &   +9.34 \\
$\Delta E_{\text{Pad\'e}}(3/3)$    & +0.047 &  +0.761 &  -1.32 &  -2.34 &    -6.00 &  -16.04 \\
$\Delta E_{\text{Pad\'e}}(3/4)$    & +0.008 &  +0.066 &  +0.66 &  +1.48 &    +2.34 &   +4.01 \\
$\Delta E_{\text{Pad\'e}}(4/3)$    & +0.132 &  +2.666 &  -0.11 &  -0.11 &    -0.88 &   -2.96 \\
$\Delta E_{\text{Pad\'e}}(4/4)$    & +0.019 &  +0.492 &  +0.18 &  +0.05 &    +3.45 &  +26.11 \\
$\Delta E_{\text{Pad\'e}}(4/5)$    & +0.015 &  +0.314 &  -0.80 &  +0.03 &    +2.43 &   +6.46 \\
$\Delta E_{\text{Pad\'e}}(5/4)$    & +0.012 &  +0.135 &  +0.03 &  +0.03 &    +0.22 &   +1.14 \\ 
$\Delta E_{\text{Pad\'e}}(5/5)$    & -0.029 &  +0.136 &  -0.04 &  +0.05 &    -0.50 &   -2.45 \\
$\Delta E_{\text{Pad\'e}}(6/6)$    & -0.037 &  +0.040 &  +0.01 &  +0.00 &    +0.16 &   +1.37 \\ 
$\Delta E_{\text{Pad\'e}}(8/8)$    & +0.001 &  +0.001 &  +0.01 &  +0.01 &    +0.01 &   +0.20 \\
$\Delta E_{\text{Pad\'e}}(10/10)$  & +0.000 &  +0.001 &  +0.01 &  +0.01 &    +0.00 &   +0.01 \\
$\Delta E_{\text{Pad\'e}}(12/12)$  & -0.000 &  -0.002 &  +0.01 &  +0.01 &    +0.00 &   +0.01 \\
$\Delta E_{\text{Pad\'e}}(15/15)$  & -0.000 &  -0.000 &  +0.01 &  +0.01 &    +0.01 &   -0.04 \\
\hline
\end{tabular}
\end{table}

A quantitative comparison is presented in Tab.~\ref{tab:comparison}, where the difference of the partial sums $E_{\text{sum}}(p)$ and the Pad\'e approximants $E_{\text{Pad\'e}}(M/N)$ to the exact NCSM eigenvalues $E_{\text{NCSM}}$ are shown. The latter were obtained using the \textsc{Antoine} shell-model code \cite{CaNo99}. Starting from $M+N\approx10$ the deviations of the Pad\'e approximants from the exact result are getting very small and starting from $M+N\approx20$ the Pad\'e approximants are numerically identical to the exact result for all nuclei. In this regime the individual MBPT contributions $E^{(p)}$ are already increasing exponentially for \elem{O}{16} and \elem{Ca}{40} (cf. Fig.~\ref{fig:series}) and the partial sum  $E_{\text{sum}}(p)$ does not provide any sensible estimate of the ground-state energy. The Pad\'e approximants prove to be a highly efficient tool to extract a virtually exact and stable result for the energy from the first 10 or more coefficients $E^{(p)}$ of the strongly fluctuating and non-converging power series. Considering the scale of the order-to-order fluctuations and the absolute size of the perturbative contributions $E^{(p)}$ the stability and the precision of the converged Pad\'e approximants is truly remarkable.
 
For application purposes, the behavior at low orders is also of interest. As shown in Tab.~\ref{tab:comparison}, the deviations $\Delta E_{\text{sum}}(p)$ and $\Delta E_{\text{Pad\'e}}(M/N)$ are of comparable magnitude up to about fifth order, both showing sizable fluctuations. Hence, in this low-order domain the Pad\'e approximants do not improve on the results obtained from a simple partial sum. Only beyond $M+N\approx5$ do the Pad\'e approximants start to converge, i.e., the variations within a set of approximants of neighboring order reduce systematically. At the same time the deviations of the partial sums, $\Delta E_{\text{sum}}(p)$, start to increase exponentially. 

The stability of the Pad\'e approximants $E_{\text{Pad\'e}}(M/N)$ across various neighboring orders $M$ and $N$ is an important intrinsic criterion for convergence and for the accuracy of the Pad\'e approximants as compared to the exact result. Therefore, it seems advisable to always consider sets of several approximants. Moreover, there is always the possibility that individual approximants completely escape the general trend, such as the $E_{\text{Pad\'e}}(2/2)$ approximant for \elem{Ca}{40} at $\hbar\Omega=20$ MeV that has a large positive and thus unphysical value. These cases are a reminder that the convergence theorems for Pad\'e approximants, e.g., the Pad\'e conjecture, only cover subsequences of approximants. Finally we note that the MBPT power series in the present examples turns out not to be a Stieltjes series. As the Pad\'e approximants of Tab.~\ref{tab:comparison} show, the inequality \eqref{eq:inequality} as well as related inequalities are not fulfilled. The Pad\'e approximants for the ground state energy on nuclei in the present MBPT framework do not provide rigorous bounds for the exact eigenvalues.

\section{Conclusions \& Outlook}

We have formulated and applied many-body perturbation theory up to high orders for the description of ground-state energies of closed-shell nuclei using realistic Hamiltonians. In contrast to typical applications of MBPT in nuclear physics that are limited to second or third order, we extend the order-by-order evaluation of the perturbation series up to 30th order using a simple recursive scheme. In order to facilitate the comparison with exact eigenvalues obtained in NCSM calculations, we have limited ourselves to an harmonic-oscillator single-particle basis and a $N_{\max}\hbar\Omega$ space. However, results for other single-particle bases and model-space truncations are qualitatively similar.

Our major results and conclusions are: First, a simple partial sum of the perturbative series in general does not converge. On the contrary, we typically observe an exponential increase of the individual perturbative contributions $|E^{(p)}|$ and an oscillatory behavior of the partial sum $E_{\text{sum}}(p)$ as function of $p$. Thus, finite partial summations, even if they are extended to high orders, do not provide a stable and systematically improvable approximation for the exact energy eigenvalue. 

Second, Pad\'e approximants offer a computationally simple yet powerful tool to extract a convergent series from a finite set of perturbative energy contributions $E^{(p)}$. Solely through a resummation of the finite $p$th-order power series to a rational function, whose Taylor expansion up to order $p$ is identical to the initial power series, we are able to extract a highly stable and convergent approximation for the energy. The information entering these Pad\'e approximants of order $p=M+N$ is identical to the power series of order $p$ and so is the computational effort. However, whereas a finite partial sum $E_{\text{sum}}(p)$ explodes with increasing order $p$, the Pad\'e approximants $E_{\text{Pad\'e}}(M/N)$ converge.      

Third, beyond a sufficiently large order, typically $M+N\gtrsim10$, the different Pad\'e approximants become very stable and converge rapidly to a unique value for the energy. This energy is identical to the exact eigenvalue obtained in the corresponding NCSM calculation. In this sense, Pad\'e resummed MBPT and the Lanczos diagonalization become numerically equivalent. Unfortunately, the computational effort is also comparable at least for the implementation we adopted for the MBPT here.

Fourth, at low orders, i.e. $M+N\lesssim4$, the Pad\'e approximants generally do not yield an improved approximation for the energy. Their deviations from the exact eigenvalue fluctuate and are of the same order of magnitude as the errors of the corresponding partial sums. If one is limited, for computational reasons, to very low orders, then the version of MBPT used here can only provide a rough estimate that might differ significantly from the exact eigenvalue.

These initial studies open a number of new avenues for the study and application of MBPT in nuclear structure. Beyond the MBPT calculations presented here, one can use optimized single-particle bases and improved partitionings of the Hamiltonian to influence the convergence behavior of a finite order-by-order MBPT calculation. Furthermore, one can exploit infinite partial summations, e.g., ladder- or ring-type summations including MBPT contributions from all orders, and compare their quality to Pad\'e resummed finite-order calculations. These improvements and extensions will be the subject of future studies.

\section*{Acknowledgments}

This work is supported by the DFG through contract SFB 634, the Helmholtz International Center for FAIR within the framework of the LOEWE program launched by the State of Hesse, and the BMBF through contract 06DA9040I.


\end{document}